\documentclass[9pt,twocolumn,twoside]{IEEEtran}

\usepackage{amsmath,amsfonts}
\usepackage{algorithmic}
\usepackage{algorithm}
\usepackage{array}
\usepackage[caption=false,font=normalsize,labelfont=sf,textfont=sf]{subfig}
\usepackage{textcomp}
\usepackage{stfloats}
\usepackage{url}
\usepackage{verbatim}
\usepackage{graphicx}
\usepackage{cite}
\begin{document}
	
	\title{Quantum-Key Distribution using Decoy Pulses to Combat Photon-Number Splitting by Eavesdropper: An Event-by-Event Impairment Enumeration Approach for Performance Evaluation and Design}
	
	\author{Debasish Datta,~\IEEEmembership{Life Senior Member,~IEEE}
}
	
	
	
	\maketitle








\begin{abstract}
	Quantum-key distribution (QKD) schemes employing quantum communication links are typically based on the transmission of weak optical pulses over optical fibers to setup a secret key between the transmitting and receiving nodes. The sender (Alice) transmits optically a random bit stream to the receiver (Bob) through the photon polarizations or the quadrature components of the lightwaves associated with the photons, with a secret key remaining implicitly embedded therein. However, during the above transmission, some eavesdropper (Eve) might attempt to tap the passing-by photons from the optical fiber links to extract the key. In one of the popular QKD schemes, along with signal pulses, some additional decoy pulses are transmitted by Alice, while Eve might use photon-number splitting (PNS) for eavesdropping. In a typical PNS scheme, (i) the optical pulses with single photon are blocked by Eve, (ii) from the optical pulses with two photons, one photon is retained by Eve to carry out eavesdropping operation and the other is retransmitted to Bob, and (iii) all other pulses with more than two photons are retransmitted by Eve to Bob without retaining any photon from them. Extensive theoretical research has been carried out on such QKD schemes, mostly by employing information-theoretic approach along with computer simulations and experimental studies. In this paper, we present a novel event-by-event impairment enumeration approach to evaluate the overall performance of one such QKD scheme analytically with due consideration to the physical layer of the quantum communication links. The proposed approach monitors the impairments of the propagating optical pulses event-by-event at all possible locations along the optical fiber link using statistical approach, and provides estimates of the realizable key generation rate, while assuring an adequate yield ratio between signal and decoy pulses for the detection of possible eavesdropping.
	
\end{abstract}

%

\section{Introduction}
\label{intro}
Various underlying transmission impairments prevalent in quantum communication systems deserve critical attention while designing a secure communication network employing quantum-key distribution (QKD). Robust QKD schemes generally employ quantum communication through dedicated optical fiber links (quantum channels) to setup a secret key between the transmitting and receiving nodes. Typically, the sender (Alice) transmits over an optical fiber link a random bit stream to the receiver (Bob) through photon polarizations or encoded quadrature components of the lightwaves associated with photons, leading to two basic categories of QKD schemes, viz, discrete-variable QKD (DV-QKD) and continuous variable QKD (CV-QKD), respectively. In a DV-QKD scheme, Alice sends Bob a train of optical pulses, preferably with single photon, whose polarizations are encoded with random data stream. While receiving the optical pulses from Alice, Bob chooses his polarization setting randomly for each optical pulse, causing thereby unavoidable photon losses. Thereafter, the data retrieved by Bob is used by him as well as Alice collaboratively through public channel to distil a secret quantum key (or, simply secret key). In a CV-QKD scheme, Alice sends light by encoding the quadrature components of the lightwave associated with the emitted photons, which are received by Bob by using coherent (homodyne or heterodyne) detection for data retrieval. Thereafter, the received data bits, as in DV-QKD, are used by Bob and Alice collaboratively to process and distil a secret key. In such QKD systems, typically an eavesdropper (Eve) engages herself somewhere between Alice and Bob, and attempts to tap the photons transmitted by Alice through the optical fiber used for the quantum channel. Following the pioneering work of Bennett and Brassard on QKD in 1984, viz., BB84 protocol \cite{Ref1}, there have been vigorous investigations on QKD schemes during the last four decades \cite{Ref2}, \cite{Ref3}, \cite{Ref4}, \cite{Ref5}, \cite{Ref6}, \cite{Ref7}, \cite{Ref8}. In the present work, considering the availability of more mature technology for DV-QKD schemes, we consider a specific DV-QKD scheme for further investigation and attempt to develop a practical design methodology for the same \cite{Ref9}, \cite{Ref10}, \cite{Ref11}.

In the DV-QKD schemes, Eve can choose different possible mechanisms to eavesdrop into the quantum channel, including the basic BB84 scheme and various other versions of the same. In one of these BB84-based schemes, she can simply use a beam-splitter to tap the passing-by optical pulses carrying one or more photon(s). In a more aggressive scheme, Eve can intercept the link and take out the entire optical pulse which might carry one or multiple photon(s). Thereafter she can employ a strategy, known as photon-number splitting (PNS) scheme, wherein if she receives a single-photon optical pulse, she blocks the photon so that it doesn't reach Bob. Thus, Eve passes on the zero-photon slots, blocks all the single-photon pulses, and retransmits the multi-photon pulses to Bob. For the two-photon pulses, Eve retains one and retransmits the other to Bob, while she retransmits to Bob all photons for the remaining multi-photon pulses with three or more photons. In order to combat with the PNS attack from Eve, one intelligent scheme has been studied using the transmission of optical pulses having different intensity levels, e.g., signal and decoy pulses, wherein the decoy pulses are inserted randomly into the signal pulse stream for transmission \cite{Ref12}, \cite{Ref13}. Bob with the help of Alice through public channel can identify the decoy pulses and thus can detect the presence of Eve, if any, by differentiating the \textit{yields} of signal and decoy pulses (estimates of the probabilities of successful reception of signal and decoy pulses) at Bob's receiver. Further, the reception of the decoy pulses (along with the signal pulses) makes it difficult for Eve to decipher the right photons carrying the desired signal, causing hurdles for her in extracting the secret code. The protocol used in this DV-QKD scheme (hereafter, simply QKD scheme) is thus an extended version of the BB84 scheme using decoy pulses against PNS attack from Eve, and is hereafter abbreviated as BB84-DP-PNS protocol. We consider this protocol in the following for the present study.

In all QKD schemes, while working against eavesdropping, the overall design becomes more challenging due to the presence of various transmission impairments in the quantum channel. One of the potential transmission impairments (henceforth, simply impairments) is encountered from the possible losses of photons in optical fiber links used in the quantum channel, which would be ideally based on the transmission of single-photon optical pulses. While the loss of optical power in optical fibers can be measured using standard techniques, any method to estimate the probability of loss of a \textit{single photon} has not been modelled explicitly so far. Photon losses are also unavoidable due to possible mismatch between the polarizations of the transmitted photons and the polarization-settings at the receivers. Furthermore, in practice single-photon emission from an optical source in a given time slot cannot be guaranteed due to the inherent non-deterministic (Poissonian) nature of photon emission process in optical sources. Even from the photons that survive through the lossy fiber and polarization mismatch, the receivers might at times fail to extract the needed information due to the impairments encountered in the photodetection process and subsequent signal processing stages in presence of omnipresent noise contamination. 

To evaluate the performance for BB84-DP-PNS protocol in the light of information theory, one needs to find out the key generation rate between Alice and Bob using the estimates of entropies at Bob's receiver \cite{Ref13}, \cite{Ref14}, \cite{Ref15}, \cite{Ref16}. However, in the present work, we obtain similar estimate of the key generation rate, but through a novel event-by-event enumeration of the impairments of optical pulses at all the relevant locations along the quantum channel from Alice to Bob through Eve. In particular, we walk alongside the propagating optical pulses through the quantum channel to capture event-by-event all the impairments at different locations (viz., at Alice's transmitter, optical fibers, Eve's receiving and retransmitting operations for implementing PNS attack, and Bob's receiver) to develop a comprehensive impairment enumeration model and use the same to evaluate the realizable key generation rate, along with the estimates of the signal-to-decoy yield ratio observed by Bob and the signal-to-decoy pulse reception ratio in Eve's receiver for facilitating the overall system design.

The rest of the paper is organized as follows. Section 2 presents the basic operation of BB84-DP-PNS protocol. In Section 3, some of the potential impairment phenomena are first examined independently and combined thereafter with a few others to obtain an event-by-event comprehensive model to enumerate and capture all of them in terms of the probabilities of successful reception of photons and the numbers of signal and decoy bits retrieved by Bob and Eve. Subsequently, in Section 4 the above model is used to obtain the estimates of (i) key generation rate between Alice and Bob, (ii) signal-to-decoy yield ratio observed by Bob, and (iii) signal-to-decoy pulse reception ratio in Eve's receiver. Section 5 presents the results obtained by using the proposed model for  the BB84-DP-PNS protocol, and the implications of the results are discussed in the context of overall system design. Finally, Section 6 concludes the paper. 

\section{DV-QKD Scheme using BB84-DP-PNS Protocol}
\label{sec:1}

\noindent Figure 1 shows a block schematic of BB84-DP-PNS protocol, wherein Alice transmits for Bob a random binary stream over the quantum channel by using photon polarizations, which Eve intercepts (without the knowledge of Alice and Bob) and applies PNS scheme and retransmits photons selectively to Bob. Alice uses a stream of $m = m_s + m_d + m_v$ synchronized time slots to generate current pulses for driving her transmitting laser with signal, decoy, and vacuum pulses, out of which $m_s$ slots carry signal pulses, $m_d$ slots carry decoy pulses, and $m_v$ slots carry no photon, i.e., vacuum pulses (for the measurement of dark current effects), which are mixed (i.e., multiplexed) randomly over time. All the signal slots carry some binary information (bits) represented by the polarizations of the constituent photons, which may be one or more in a given optical pulse during a time slot, while the polarization associated with the decoy pulses are of no consequence to Bob. Note that, in a given optical pulse carrying multiple photons, all the photons remain polarized identically, and as indicated earlier the polarization of a received photon may or may not match with the receiver polarization-setting of Bob/Eve.

In order to employ eavesdropping, Eve intercepts the fiber link between Alice and Bob and receives all the optical pulses from Alice, with zero, one or more photons for both signal and decoy pulses (along with vacuum pulses), presuming that the direct fiber link between Alice and Bob has been stealthily intercepted by Eve with alternate fiber link between Eve and Bob at an opportune time and location. Eve uses PNS scheme on the intercepted optical pulses, using the functional blocks, Eve-1 and Eve-2, as shown in Fig.1, for fiber interception and photon reception followed by the decoding needed for extracting the secret key and selective retransmission of photons towards Bob. Next, Bob receives the optical pulses from Eve, takes note of all the received vacuum, decoy and the signal pulses. Vacuum pulses are used to assess the effect of dark current in the receiver, while both signal and decoy pulses are counted for estimating the respective probabilities of successful pulse reception leading to the estimates of the signal and decoy yields. The signal pulses are thereafter subjected to \textit{sifting}, photodetection, bit detection from the photodetected signal, and forward error-correction (FEC) \cite{Ref1}, eventually leading to a shorter bit stream. The secret key is finally \textit{distilled} out from this sifted and error-corrected signal bit stream by using an appropriate privacy amplification (PA) scheme \cite{Ref14}, which uses a publicly-announced compression algorithm. Next, we discuss how the various steps of the QKD protocol are carried out by Alice and Bob in further details.

\begin{figure}[ht]
\centering
\includegraphics[width = 4in, viewport = 2 4 790 530]{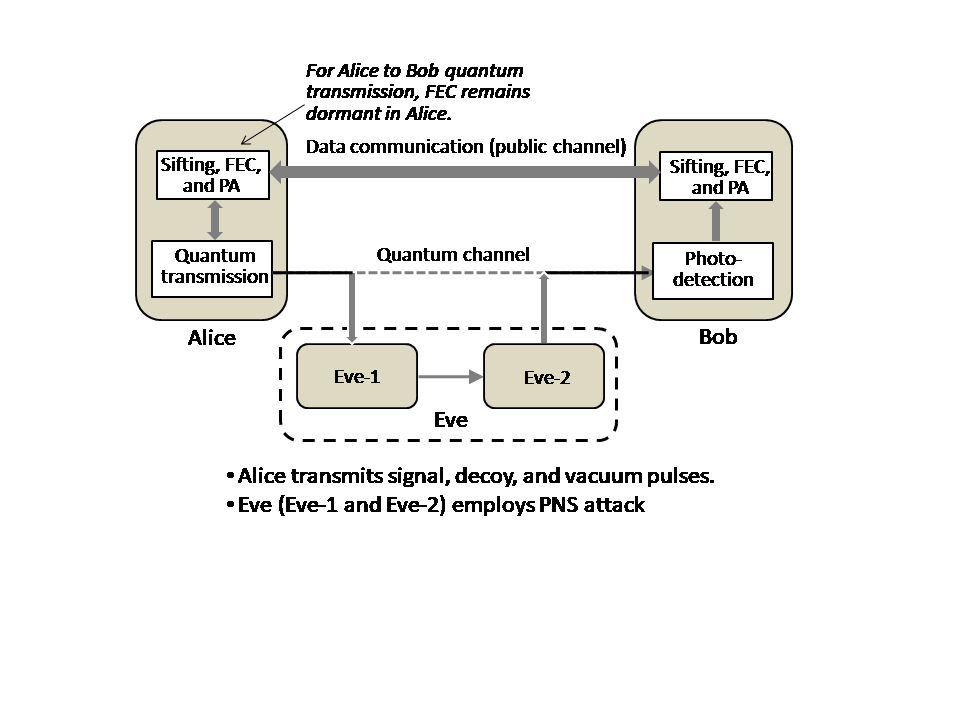}
\caption{Block schematic of QKD system using QKD-PD-PNS protocol.}
\end{figure}

For each optical pulse carrying photon(s), the photon polarization basis is randomly chosen by Alice to be either a rectilinear (R) orientation (basis) with vertical or horizontal polarization for binary 1 and 0, respectively, or a diagonal (D) orientation with $45^\circ$ or $135^\circ$ polarization for binary 1 and 0, respectively. In the background, a synchronization subsystem operates in public channel to realize slot-synchronized (i.e., clocked) reception of optical pulses.

At the receiving end, for each time slot, Bob (and Eve too) chooses randomly either R or D base at its receiver (regardless of what basis Alice used) and hence on the average both fail to receive half of Alice's/Eve's photons. Nevertheless, for rest of the photons, Bob (and Eve too) receives with the correct polarizations of the received photons and tries to retrieve the respective binary 1's and 0's, after photodetection. 

In particular, after the reception of optical pulses, Bob would collaboratively work with Alice on public channel to perform a number of functions on the retrieved bits, viz., sifting, photodetection and bit detection, FEC, and compression of the error-corrected bit stream by using appropriate PA algorithm. For carrying out the sifting operation, first Bob declares its polarization bases (R or D) used for each signal pulse, following which Alice lets Bob know what bases she used for those pulses. Thereafter, Bob reveals some of the bits (extracted from the polarizations of the successfully received optical pulses) to Alice in public channel which are subsequently confirmed by Alice, and these bits are not used in the key formation process. Next, they interact on public channel for FEC, by exchanging parity bits of smaller segments of the bit streams, which also leads to some reduction in the number of available bits. Thus, following sifting and FEC operations, Bob settles down with much fewer bits as compared to $m$. Finally, Alice and Bob form a yet shorter string of bits by executing a suitable PA algorithm on the sifted and error-corrected bits using some publicly-announced compression function, thereby leading to the formation of the desired key, also known as quantum key, which eventually ensures security for the associated network.

\section{Impairments in the BB84-DP-PNS Protocol}

\noindent In this section we identify and model the relevant impairment phenomena at the appropriate locations in the QKD protocol under consideration, viz., probability of photon losses in optical fiber, polarization mismatch between the received photon and the receiver setting, uncertainties in photon transmission from optical sources, limited photodetection efficiency in receivers, and the data losses in sifting and FEC operations carried out by  Bob and Alice collaboratively.

\subsection{Probability of Photon Losses in Optical Fiber}
\label{plb}

\noindent Photon arrival process in an optical receiver during a given time slot is determined by the photon emission process in the optical source used in transmitter during the same time slot and the photon losses incurred during the course of photon propagation in the optical fiber. The photon emission process during a time slot in an optical source, such as laser, is governed by Poisson distribution, for a given average number $\lambda$ of photons in a time slot, and $\lambda$ is in turn proportional to the average transmitted optical power $P_T$ during the same time slot. For a given $\lambda$ corresponding to a given $P_T$, one can measure the received optical power $P_R$, which is also proportional to the average number $n$ of photons arrived at the receiver during the corresponding time slot. Thus, the optical power loss incurred in the optical fiber link can be expressed as the ratio $\rho$, given by

\begin{equation}
\rho = P_T/P_R = \lambda/n,
\end{equation}

\noindent Further, the fiber loss $L_{dB}$ in dB and the fiber loss $\rho$ can be expressed as

\begin{equation}
L_{dB} = \alpha l = 10\log_{10} \rho,
\end{equation}

\noindent where $\alpha$ represents the fiber loss in dB/km and $l$ is the fiber length in km, leading to

\begin{equation}
\rho = 10^{(\alpha l/10)} = 10^{(L_{dB}/10)}.
\end{equation}

\noindent As mentioned before, the photon emission statistics in the transmitter follows Poisson distribution, given by $P_{\lambda}(\mu)$ with an average photon count $\lambda$ per time slot. Similarly, the photon arrival statistics in the receiver would follow Poisson distribution $P_n(\nu)$ with an average received photon count $n$ per time slot. Hence, $P_{\lambda}(\mu)$ and $P_n(\nu)$ can be expressed as \cite{Ref17}

\begin{equation}
P_{\lambda}(\mu) = \frac{\lambda^{\mu} \exp(-\lambda)}{\mu!},
\end{equation}

\begin{equation}
P_n(\nu) = \frac{n^{\nu} \exp(-n)}{\nu!}.
\end{equation}

\noindent Using the above expressions, we proceed to develop an analytical model to evaluate the probability $p_{fl}$ of loss for one single photon in an optical fiber. In order to estimate $p_{fl}$, we consider a fiber optic link wherein a few photons ($\mu$ photons) during a given time slot are transmitted and the receiver receives $\nu (\le \mu)$ photons in the same time slot, implying that in the optical fiber link $\kappa = \mu - \nu$ photons have been lost in that time slot.

Next, we note that $\nu$ photons are received at the fiber end with a loss ratio $\mu/\nu (\ge 1)$, and obtain the loss probability $q_{\kappa}$ for $\kappa = \mu - \nu$ photons in terms of $p_{fl}$ as

\begin{equation}
q_{\kappa} = \binom{\mu}{\kappa} p_{fl}^{\kappa} (1-p_{fl})^{\mu-\kappa} = \binom{\mu}{\kappa} p_{fl}^{\kappa} (1-p_{fl})^{\nu},
\end{equation}

\noindent wherein we assume that the losses of $\kappa$ photons are all independent of each other. Therefore, one can express $n$ using $q_{\kappa}$ and $\lambda$ as

\begin{eqnarray}
n & = & \lambda - E_{\mu}\bigg\{\sum_{\kappa = 0}^{\mu}\kappa q_{\kappa}\bigg\}\\ \nonumber
  & = & \lambda - \sum_{\mu = 0}^{\infty} P_{\lambda}(\mu) \bigg\{\sum_{\kappa=0}^{\mu}\kappa \binom{\mu}{\kappa} p_{fl}^{\kappa} (1-p_{fl})^{\mu-\kappa}\bigg\}
\end{eqnarray}

\noindent Solving numerically the above expression, one can determine the value of $p_{fl}$ for an optical fiber with the given values of $\lambda$ and $n$ ( = $\lambda/\rho$) with $\rho$ being related to the fiber length $l$ (in km) and the fiber loss $\alpha$ (in dB/km) through (3).

The formulation in (7) has been used to determine the values of $p_{fl}$ for three example cases with $\alpha$ = 0.2 dB/km, viz., $l = 15$ km, $50$ km, and $100$ km. In all the three cases we considered $\lambda = 100$, leading to the values of $\rho$ as 2, 10, and 100 for the three cases, respectively. The results for $p_{fl}$ obtained through numerical computation are presented in Table 1.

\begin{table}[ht]
\begin{center}
\caption{Photon loss probability in optical fibers with $\alpha = 0.2$ dB/km.}
\begin{tabular}{c c c c c c}
\hline
$l$ in km & $L_{dB}$ & $\rho$ & $\lambda$ & $n$ & $p_{fl}$\\ \hline
15 & 3  & 2  & 100 & 50 & 0.50 \\ \hline
50 & 10 & 10 & 100 & 10 & 0.90\\ \hline
100 & 20 & 100 & 100 & 1 & 0.01\\ \hline
\end{tabular}
\end{center}
\label{tab:01}
\end{table}

A close look into the results in Table 1 indicates that in each case $p_{fl}$ assumes a value determined by $\lambda$ and $n$, given by

\begin{equation}
p_{fl} = \frac{\lambda-n}{\lambda} = 1 - \frac{1}{\rho}
\end{equation}

\noindent which was expected intuitively, as all the photon losses occur independently.

\subsection{Uncertainty in Single-Photon Transmission}

\noindent Although single-photon transmission is most desirable from Alice, in practice it is hard to ensure the same. At the best, in each time slot one can ensure an optical pulse from the source with a low value for $\lambda$, typically remaining below one. This implies that, in each time slot assigned for an optical pulse, there can be several possibilities in terms of the number $j$ of emitted photons, such as, $j = 0, 1, 2, 3$ or more with the respective probabilities, $\phi_j$'s, as governed by the Poisson distribution $P_\lambda(j) = \lambda^j \exp(-\lambda)/j!$. Thus, we obtain the emission probabilities as

\begin{eqnarray}
\phi_0 & = & P_\lambda(0) = \exp(-\lambda) \\
\phi_1 & = & P_\lambda(1) = \lambda \exp(-\lambda) \\
\phi_2 & = & P_\lambda(2) = \lambda^2 \exp(-\lambda)/2! \\
\phi_3 & = & P_\lambda(3) = \lambda^3 \exp(-\lambda)/3! \\
\phi_4 & = & P_\lambda(4) = \lambda^4 \exp(-\lambda)/4!,
\end{eqnarray}

\noindent and so on. For example, with $\lambda = 0.5$, we get $\phi_0 = 0.6065$, $\phi_1 = 0.3033$, $\phi_2 = 0.0758$, $\phi_3 = 0.0126$ and so on. This implies that, with $\lambda = 0.5$, one has to bear with about $61\%$ time slots going empty, and the rest of the time slots will carry one or more photons, with the probability of four photons being as small as $0.0016$. We shall get back to this aspect later in further details.

\subsection{Polarization Mismatch, Photodetection, Sifting, and Error Correction}

\noindent In this section, we consider the remaining impairments and discuss how they would influence the receptions of photons at the receivers of Bob and Eve. In regard to the polarization match/mismatch between the received photon and Bob's/Eve's receiver settings, we note that all the recipients keep changing their polarization settings randomly and independently. Thus, one can assume that on the average half of the transmitted photons will be lost due to the polarization mismatch both for Bob and Eve, leading to a probability of polarization loss $p_{pl} = 0.5$. Moreover, during the sifting process, while validating the correctness of the data received by Bob, some information would be exchanged by Bob and Alice on public channel. Consequently, some of the data received by Bob would get exposed to Eve and hence would be removed, leading to sifting losses \cite{Ref18}, represented by the sifting-loss factor $\alpha_{sift}$. Further, the photodetector would incur photon losses with a limited quantum efficiency $\eta_{pd}$ due to the absorption process therein \cite{Ref18}, \cite{Ref19}, followed by the errors in bit detection and subsequent FEC operation causing some more bit losses \cite{Ref18}, represented by a loss factor $\alpha_{err}$.

\subsection{Event-by-Event Enumeration of Impairments along the Optical Path from Alice to Bob through Eve}

\noindent For enumeration of all the impairments, we note that, Eve applies PNS attack with appropriate devices by intercepting the optical fiber link between Alice and Bob (stealthily, as mentioned earlier), and thus Bob does not get to receive any photon directly from Eve. In a given time slot, Alice would  transmit (besides vacuum slots) zero, one, or more photons with different probabilities, be it signal or decoy slots, and all of the transmitted photons may or may not succeed to reach Eve having gone through the lossy fiber and polarization mismatch at the receiver of Eve. When Bob receives the photons in a time slot via Eve, he tries to extract the needed information from the photons. In order to carry out the enumeration process, we start from Alice. In particular, we consider the transmission probabilities of a photon from Alice for signal and decoy pulses. Following Poissonian statistics, the probability that $j$ (with $j = 0, 1, 2, \cdot \cdot \cdot, \infty$) photons are transmitted by Alice in a time slot for a signal pulse, is expressed as

\begin{equation}
p_{tx}^{as}(j) = \frac{\lambda_s^je^{-\lambda_s}}{j!},
\end{equation}

\noindent where $\lambda_s$ represents the average photon count in the transmitted signal pulse. Similarly, the probability that $j$ photons are transmitted by Alice in a time slot for a decoy pulse, is expressed as

\begin{equation}
p_{tx}^{ad}(j) = \frac{\lambda_d^je^{-\lambda_d}}{j!},
\end{equation}

\noindent with $\lambda_d$ representing the average photon count in the transmitted decoy pulses. Further, we recall that a photon once emitted by Alice would reach Eve and be received with correct polarization, if $(i)$ the photon suffers no loss in optical fiber and $(ii)$ the polarization bases of Alice and Eve don't differ. This can occur with a probability of successful reception in optical domain at Eve's receiver, given by

\begin{equation}
\psi_{ae} = (1-p_{fl,ae})(1-p_{pl}),
\end{equation}

\noindent where $p_{fl,ae} = 1 - 1/\rho_{ae}$ represents the probability of photon loss in the optical fiber link between Alice and Eve, with $\rho_{ae}$ as the optical power loss ratio (see (8)) in the fiber segment between Alice and Eve.

Using the above definitions, we next introduce a model to represent the impact of all the above impairments event-by-event, as observed at the receivers of Eve and Bob. First, we consider the various possible scenarios that would be encountered by Eve, while receiving optical pulses from Eve. We recall that, having received the optical pulses from Alice, Eve would use her PNS scheme to transmit photons selectively in each optical pulse towards Bob.

To start with, we concentrate on a time slot, during which zero, one, or more photons (i.e., $j = 0,1, 2, 3, ...$) have been transmitted by Alice for signal and decoy pulses, and estimate the probabilities of receiving these photons successfully at Eve's receiver in optical domain. Since Eve blocks all pulses with one photon, first we consider the probabilities of reception of two photons in a time slot by Eve from Alice for signal and decoy pulses, $p_{rx,e}^{as}(2)$ and $p_{rx,e}^{as}(2)$ respectively, which are expressed as

\begin{eqnarray}
p_{rx}^{es}(2) & = & p_{tx}^{as}(2)\psi_{ae}^2 + p_{tx}^{as}(3)[3\psi_{ae}^2(1 - \psi_{ae})] + \\ \nonumber
               &   & p_{tx}^{as}(4)[6\psi_{ae}^2(1 - \psi_{ae})^2],
\end{eqnarray}

\begin{eqnarray}
p_{rx}^{ed}(2) & = & p_{tx}^{ad}(2)\psi_{ae}^2 + p_{tx}^{ad}(3)[3\psi_{ae}^2(1 - \psi_{ae})] + \\ \nonumber
               &   & p_{tx}^{ad}(4)[6\psi_{ae}^2(1 - \psi_{ae})^2],
\end{eqnarray}

\noindent where the first term in each probability is the product of two factors: $(i)$ probability of transmission of two photons per slot from Alice, and $(ii)$ probability of successful reception of both photons in optical domain at Eve. The second and third terms for each probability are added to account for the cases where three and four photons are transmitted by Alice with one and two photons being lost on the way, respectively (transmissions of five or more photons per pulse are ignored for the values of $\lambda_s$ and $\lambda_d$ to be chosen later in the case study). 

Using equations (17) and (18), and ignoring the probabilities of the optical pulses transmitted with five or more photons, we can now express the total number $n^e(2)$ of two-photon pulses received by Eve from Alice as

\begin{equation}
n^e(2) = \underbrace{m_s p_{rx}^{es}(2)}_{n^{es}(2)} + \underbrace{m_d p_{rx}^{ed}(2)}_{n^{ed}(2)},
\end{equation}

\noindent with the first term $n^{es}(2)$ representing the contribution from signal pulses and the second term $n^{ed}(2)$ contributed by decoy pulses. Note that, according to the PNS scheme as discussed earlier, having received these $n^e(2)$ pulses, each carrying two photons, Eve retains one of them and retransmits the other photon to Bob. For all other received optical pulses having three or more photons, Eve forwards all of them to Bob without retaining any photon from them. 

Next, down the optical fiber link, we consider the reception process at Bob's receiver from Eve. The probability of successful photon reception by Bob (from Eve) due to photon loss in the fiber link between Eve and Bob and the possible mismatch of polarization settings is obtained as

\begin{equation}
\psi_{eb} = (1 - p_{fl,eb})(1 - p_{pl}).
\end{equation}

\noindent where $p_{fl,eb} = 1 - 1/\rho_{eb}$ represents the probability of photon loss in the optical fiber link between Eve and Bob, with $\rho_{eb}$ as the optical power loss in the fiber segment between Eve and Bob. As mentioned earlier, Eve blocks all one-photon pulses, retains one photon from the two-photon optical pulses and forwards all photons for the optical pulses with three or more photons. Keeping this scheme in mind, we express the number of optical pulses from signal pulse stream received by Bob as

\begin{equation}
n^{bs}_{op} = n^{bs}_{op}(1) + n^{bs}_{op}(2) + n^{bs}_{op}(3),
\end{equation}

\noindent where $n^{bs}_{op}(1)$, $n^{bs}_{op}(2)$, and $n^{bs}_{op}(3)$ are given by

\begin{eqnarray}
n^{bs}_{op}(1) & = & m_s[p_{rx}^{es}(2) \psi_{eb}] \\
n^{bs}_{op}(2) & = & m_s[p_{rx}^{es}(3)\{\psi_{eb}^3 + 3\psi_{eb}^2(1 - \psi_{eb}) + \\ \nonumber 
          &   & 3\psi_{eb}(1 - \psi_{eb})^2\}] \\ 
n^{bs}_{op}(3) & = & m_s[p_{rx}^{es}(4)\psi_{eb}^4 + 4p_{rx}^{es}(4)\{\psi_{eb}^3 (1 - \psi_{eb}) + \\ \nonumber
          &   & \psi_{eb}(1 - \psi_{eb})^3\} + 6p_{rx}^{es}(4)\psi_{eb}^2(1 - \psi_{eb})^2],
\end{eqnarray}

\noindent with the probabilities of receiving three- and four-photon signal pulses by Eve from Alice, $p_{rx}^{es}(3)$ and $p_{rx}^{es}(3)$ respectively, expressed as

\begin{eqnarray}
p_{rx}^{es}(3) & = & p_{tx}^{as}(3)\psi_{ae}^3 + 4p_{tx}^{as}(4)\psi_{ae}^3(1 - psi_{ae}),\\
p_{rx}^{es}(4) & = & p_{tx}^{as}(4)\psi_{ae}^4.
\end{eqnarray}

\noindent Note that the first term of $n^{bs}_{op}$ (i.e., $n^{bs}_{op}(1)$) represents the case when Eve receives successfully two photons in an optical pulse representing a signal bit, retains one of them, and retransmits the other photon to Bob. The second and third terms of $n^{bs}_{op}$, i.e., $n^{bs}_{op}(2)$ and $n^{bs}_{op}(3)$, consider the cases when Eve receives three- and four-photon signal pulses from Alice, respectively (optical pulses of five or more photons are ignored, as before), and retransmits them without retaining any photon.

Similarly, we express the number of optical pulses received from decoy pulse stream by Bob as

\begin{equation}
n^{bd}_{op} = n^{bd}_{op}(1) + n^{bd}_{op}(2) + n^{bd}_{op}(3),
\end{equation}

\noindent where $n^{bd}_{op}(1)$, $n^{bd}_{op}(2)$, and $n^{bd}_{op}(3)$ are given by

\begin{eqnarray}
n^{bd}_{op}(1) & = & m_d[p_{rx}^{ed}(2) \psi_{eb}] \\
n^{bd}_{op}(2) & =  & m_d[p_{rx}^{ed}(3)\{\psi_{eb}^3 + 3\psi_{eb}^2(1 - \psi_{eb}) + \\ \nonumber
          &   & 3\psi_{eb}(1 - \psi_{eb})^2\}] \\
n^{bd}_{op}(3) & = & m_d[p_{rx}^{ed}(4)\psi_{eb}^4 + 4p_{rx}^{ed}(4)\{\psi_{eb}^3 (1 - \psi_{eb}) +  \\ \nonumber 
          &   & \psi_{eb}(1 - \psi_{eb})^3\} + 6p_{rx}^{ed}(4)\psi_{eb}^2(1 - \psi_{eb})^2],
\end{eqnarray}

\noindent with the probabilities of receiving three- and four-photon signal pulses by Eve from Alice, $p_{rx}^{ed}(3)$ and $p_{rx}^{ed}(3)$ respectively, expressed as

\begin{eqnarray}
p_{rx}^{ed}(3) & = & p_{tx}^{ad}(3)\psi_{ae}^3 + 4p_{tx}^{as}(4)\psi_{ae}^3(1 - psi_{ae}), \\
p_{rx}^{ed}(4) & = & p_{tx}^{ad}(4)\psi_{ae}^4.
\end{eqnarray}

Next, we express the number of signal bits retrieved in Bob's receiver from the photodetected optical pulses as

\begin{equation}
n^{bs}_{pd} = n^{bs}_{pd}(1) + n^{bs}_{pd}(2) + n^{bs}_{pd}(3),
\end{equation}

\noindent where the three terms on the right-hand side are obtained using $\eta_{pd}$ in the expressions for $n^{bs}_{op}(1)$, $n^{bs}_{op}(2)$, and $n^{bs}_{op}(3)$ (equations (22), (23), and (24)) as

\begin{eqnarray}
n^{bs}_{pd}(1) & = & m_s[p_{rx}^{es}(2) \psi_{eb}] \eta_{pd} \\
n^{bs}_{pd}(2) & = & m_s[p_{rx}^{es}(3)\{\psi_{eb}^3\eta_{pd}^3 + 3\psi_{eb}^2(1 - \psi_{eb})\eta_{pd}^2 + \\ \nonumber 
               &   & 3\psi_{eb}(1 - \psi_{eb})^2\eta_{pd}\}] \\ 
n^{bs}_{pd}(3) & = & m_s[p_{rx}^{es}(4)\psi_{eb}^4\eta_{pd}^4 + \\ \nonumber
               &   & 4p_{rx}^{es}(4)\{\psi_{eb}^3 (1 - \psi_{eb})\eta_{pd}^3 + \psi_{eb}(1 - \psi_{eb})^3\eta_{pd}\} + \\ \nonumber
               &   & 6p_{rx}^{es}(4)\psi_{eb}^2(1 - \psi_{eb})^2\eta_{pd}^2].
\end{eqnarray}

Similarly, we express the number of decoy bits retrieved in Bob's receiver from the photodetected optical pulses as

\begin{equation}
n^{bd}_{pd} = n^{bs}_{pd}(1) + n^{bs}_{pd}(2) + n^{bs}_{pd}(3),
\end{equation}

\noindent with its three components, given by

\begin{eqnarray}
n^{bd}_{pd}(1) & = & m_d[p_{rx}^{ed}(2) \psi_{eb}] \eta_{pd} \\
n^{bd}_{pd}(2) & = & m_d[p_{rx}^{ed}(3)\{\psi_{eb}^3\eta_{pd}^3 + 3\psi_{eb}^2(1 - \psi_{eb})\eta_{pd}^2 + \\ \nonumber 
          &   & 3\psi_{eb}(1 - \psi_{eb})^2\eta_{pd}\}] \\ 
n^{bd}_{pd}(3) & = & m_s[p_{rx}^{ed}(4)\psi_{eb}^4\eta_{pd}^4 + \\ \nonumber
               &   & 4p_{rx}^{ed}(4)\{\psi_{eb}^3 (1 - \psi_{eb})\eta_{pd}^3 + \psi_{eb}(1 - \psi_{eb})^3\eta_{pd}\} + \\ \nonumber
               &   & 6p_{rx}^{ed}(4)\psi_{eb}^2(1 - \psi_{eb})^2\eta_{pd}^2].
\end{eqnarray}

The above formulations are utilized in the following to evaluate the performance of the QKD scheme using BB84-DP-PNS protocol.

\section{Key Generation Rate, Signal-to-Decoy Yield Ratio for Bob, and Signal-to-Decoy Pulse Reception Ratio for Eve}

In this section, we estimate the key generation rate for a given set of system parameters, along with the signal-to-decoy yield ratio for Bob and signal-to-decoy pulse ratio for Eve to assess whether or not the design is acceptable. First, we make use of the above model for the quantum channel in the Alice-Eve-Bob path, to estimate the key generation rate. In particular, we determine the number of signal bits retrieved by Bob from Eve's transmission. In doing so we note that, with each of the received optical pulses, sifting and FEC operations are carried out by Bob and Alice collaboratively with the consequent bit losses. In view of this, we express the number of signal bits detected successfully by Bob as

\begin{equation}
n_{err,sift}^{bs} = n^{bs}_{pd}(1 - \alpha_{err})(1 - \alpha_{sift}),
\end{equation}

\noindent where $\alpha_{err}$, $\alpha_{sift}$ are as defined earlier. Finally, $n_{err,sift}^{bs}$ leads to the key generation rate $R_k$, given by

\begin{equation}
R_k = n_{rx,sift}^{bs}/(m_s + m_d + m_v),
\end{equation}

Next, we estimate the yields, $y^{bs}$ and $y^{bd}$, seen at Bob for signal and decoy pulses, respectively. As mentioned earlier, yield of a signal or decoy pulse stream is the probability of successful reception of the same by Bob. Hence, we express the yields $y^{bs}$ and $y^{bd}$ as the estimated number of optical pulses in each case normalized with the respective number of transmitted optical pulses ($m_s$ and $m_d$) as

\begin{equation}
y^{bs} = \frac{n^{bs}_{op}}{m_s},
\end{equation}

\begin{equation}
y^{bd} = \frac{n^{bd}_{op}}{m_d},
\end{equation}

\noindent leading to the signal-to-decoy yield ratio $\rho^y_{sd}$ observed by Bob, given by

\begin{equation}
\rho^y_{sd} = \frac{y^{bs}}{y^{bd}}.
\end{equation}

On the other hand, the design also needs to check whether Eve gets some share of decoy pulses carrying two-photon pulses, implying that $n^{ed}(2)$ has some presence in $n^e(2) (= n^{es}(2) +n^{ed}(2)$), while ensuring an appropriate value for $\rho^y_{sd}$. In view of this, we define a ratio $\rho^e_{sd}$, termed as the signal-to-decoy pulse ratio at Eve's receiver for the signal and decoy pulses received from Alice (each pulse received with two photons), which is expressed as

\begin{equation}
\rho^e_{sd} = \frac{n^{es}(2)}{n^{ed}(2)}.
\end{equation}

Overall, we observe that, for the design to be acceptable, Bob needs to ensure $\rho^y_{sd} > 1$ (say 2, at least) for $\lambda_s > \lambda_d$ and for Eve $\rho^e_{sd}$ should not be too large, implying that the presence of decoy pulses should be non-negligible compared to signal pulses, making it hard for her to extract the key from the clutter of decoy bits. Using this guideline, we examine in the following section the QKD scheme employing BB84-DP-PNS protocol in presence of impairments.

\section{Results and Discussion}

First, we examine how the average number of photons emitted by Alice during a time slot, say $\lambda$ (= $\lambda_s$ or $\lambda_d$ in the present context), might affect the QKD scheme using BB84-PD-PNS protocol. In Table 2, we present four different cases for $\lambda$ = 0.1, 0.2, 0.5, and 1. The first row considers a small value of $\lambda$ = 0.1, which results in transmission of very few time slots from Alice carrying more than one photon ($\phi_{j > 1}$ = 0.0047, i.e., 0.047\%), and also many blank time slots ($\phi_0 = 0.9048$, i.e., about 90\%). Further, the probability of two-photon pulses is only ($\phi_2$ = 0.0045, implying that the number of decoy pulses would be rather small to detract Eve in her eavesdropping attempt. In the second row with $\lambda = 0.2$, number of blank slots decreases with $\phi_0 = 0.8187$, while the probability of two-photon pulses increases to $\phi_2 = 0.1637$. In the second row with $\lambda = 0.2$, probability of transmitting blank slots increases to $\phi_0 =  0.8187$, as compared to the first row with $lambda = 0.1$. On the other hand, for the third row with $\lambda$ = 0.5, Alice transmits, as compared to the first and second rows, fewer blank time slots ($\phi_0 = 0.6065$), while more of her time slots would carry optical pulses with one as well as multiple photons. In particular, with $\lambda = 0.5$, Eve will receive more two-photon optical pulses ($\phi_2 = 0.0758$) from which she will retain one photon from each pulse and forward one photon to Bob; however, she will retransmit to Bob all photons of the pulses with three or more photons. In the fourth row, with $\lambda = 1$ Alice will transmit yet fewer blank slots than the last three cases, while sending more of single- and multi-photon pulses. 

\begin{table}[ht]
\begin{center}
\caption{Values of $\phi_j$'s for different values of $\lambda$.}
\begin{tabular}{c c c c c}
\hline
$\lambda$ & $\phi_0$ & $\phi_1$ & $\phi_2$ & $\phi_{j>1}$ \\
\hline
 0.1   &  0.9048 &  0.0905  & 0.0045 & 0.0047 \\ \hline
 0.2   &  0.8187 & 0.1637   & 0.0164 & 0.0012  \\ \hline
 0.5   &  0.6065 & 0.3033   & 0.0758 & 0.0902 \\ \hline
 1.0   &  0.3679 &  0.3679  & 0.1839 & 0.2642 \\ \hline
\end{tabular}
\end{center}
\end{table}

Taking cue from the foregoing discussion, we consider next some typical cases with $\lambda_s = 0.5$, $m_s = 1,000,000$, $m_d$ = 500,000, and some values of $\lambda_d$ in the range [0.1,0.5], along with $m_v$ = 10,000 vacuum pulses (the value of $m_v$ needs to be adequate to examine the effects of dark current in the receiver, while a large value for $m_v$ will effectively reduce the overall key generation rate). Further, we assume that the length of the optical fiber link between Alice and Bob via Eve is $l = 50$ km, with Eve trying to eavesdrop using PNS attack from somewhere in between Alice and Bob (Eve being assumed to be placed halfway between Alice and Bob). Other relevant system parameters considered for the study are presented in Table 3 \cite{Ref18}, \cite{Ref19}.  

\begin{table}[ht]
\begin{center}
\caption{QKD system parameters.}
\begin{tabular}{c c c c}
\hline
\vspace{.02in}
$p_{pl}$ & $\alpha_{sift}$ & $\eta_{pd}$ &  $\alpha_{err}$ \\ \hline
0.5   &     0.2   &    0.3    &    0.2 \\ \hline
\end{tabular}
\end{center}
\end{table}

Using the above settings, we present in Table 4 the variations of $\rho^{e}_{sd}$, $\rho^y_{sd}$, and $R_k$ over a range of $\lambda_d$ (from 0.01 to 0.5) for a fixed value of $\lambda_s$ = 0.5. First, we consider the bottom row of the table. In this row with same value for $\lambda_s$ and $\lambda_d$ (= 0.50), we get $\rho^{e}_{sd} = 2$ and $\rho^y_{sd} = 1$. This gives a quick sanity check of the present model as follows. First, Eve having the same probability for receiving photons from signal and decoy pulses (since $\lambda_s = \lambda_d$ in this case), but with $m_s/m_d = 1,000,000/500,000 = 2$, she collects double number of two-photon pulses from signal pulse stream as compared to her decoy pulses with two photons, thereby making $\rho^e_{sd} = 2$. On the other hand, the yields of signal and decoy pulses at Bob's receiver become identical with $\lambda_s = \lambda_d$ and hence their ratio becomes unity. 

Further, we observe that, both $\rho^{e}_{sd}$ and $\rho^y_{sd}$ keep decreasing with the increase in $\lambda_d$, while $R_k$ remains constant, as governed by the fixed value of $\lambda_s$. However, too small values of $\lambda_d$ (e.g., 0.01, 0.05) yield rather large values for $\rho^{e}_{sd}$ and $\rho^y_{sd}$, wherein the visibility of decoy pulses to Bob becomes poor. Going down the table from the top row, we observe that the fifth row with $\lambda_d = 0.20$ (shown in bold) offers a reasonable choice as $\rho^y_{sd}$ is much larger (= 6.13) than its expected value (= 1) in absence of eavesdropping, thus making Eve's activity highly detectable to Bob. Note that, unlike in the case with $\lambda_s = \lambda_d$ in the bottom row of the table, with $\lambda_s \neq \lambda_d$ Eve's retransmissions of signal and decoy pulses would be governed by different statistics making thereby $\rho^y_{sd} \neq 1$, and more particularly $\rho^y_{sd} > 1$ with $\lambda_s > \lambda_d$. Further, $\rho^{e}_{sd}$ is also found to be 11.82, ensuring that Eve's knowledge of signal pulses will be cluttered by about 100/11.82 = 8.46\% decoy pulses scattered around randomly in Eve's received pulse stream with two photons, thereby making it hard for Eve to extract the secret key as well as making the task of PA for Alice/Bob less critical. However, one can also consider $\lambda_d = 0.3$ yielding $\rho^{e}_{sd} = 5.28$ (leading to more decoy clutter in Eve's received two-photon pulses, i.e., 100/5.28 = 18.94\% decoy pulses), albeit at the cost of reduction in the value of $\rho^y_{sd}$ to 2.71. In other words, we wish to give higher preference for the successful detection of Eve's presence than the success/failure of her eavesdropping process, and thus proceed with $\lambda_s = 0.5$ and $\lambda_d = 0.2$. 

\begin{table}[ht]
\begin{center}
\caption{Results obtained for the system parameters in Table 3, with $l$ = 50 km, $m_s$ = 1,000,000, $m_d$ = 500,000, $m_v$ = 10,000, $\lambda_s$ = 0.5.}
\begin{tabular}{c c c c}
\hline
$\lambda_d$ & $\rho^{e}_{sd}$ & $\rho^y_{sd}$ & $R_k$ \\
\hline
   0.01     & 4585.16 &   2436.84 & $6.08 \times 10^{-5}$ \\ \hline
   0.05     & 184.57  &   97.57   & $6.08 \times 10^{-5}$ \\ \hline
   0.10     & 46.51   &   24.42   & $6.08 \times 10^{-5}$ \\ \hline
   0.15     & 20.84   &   10.87   & $6.08 \times 10^{-5}$ \\ \hline
   \textbf{0.20} & 11.82 &   6.13    & $6.08 \times 10^{-5}$ \\ \hline
   0.30     & 5.35    &   2.74    & $6.08 \times 10^{-5}$ \\ \hline
   0.40     & 3.06    &   1.55    & $6.08 \times 10^{-5}$ \\ \hline
   0.50     & 2       &   1       & $6.08 \times 10^{-5}$ \\ \hline
\end{tabular}
\end{center}
\end{table}

Next, we present in Fig. 2 the variation of $R_k$  with $l$ while keeping $\lambda_s = 0.5$ and $\lambda_d = 0.2$, along with the same system parameters, wherein the plot exhibits the progressive fall (linear in log scale) of $R_k$ with increasing $l$, as expected. Further, for all the values of $l$ in Fig. 2, we also present in Fig. 3 the plots of $\rho^{e}_{sd}$ and $\rho^y_{sd}$ versus $l$ with the same system parameters, which ensures that both of them remain confined around the acceptable ranges as in the case with $l = 50$ km (i.e., in the vicinity of 11 and 6 for $\rho^{e}_{sd}$ and $\rho^y_{sd}$, respectively).

\begin{figure}[ht]
\centering
\includegraphics[width=4in, viewport= 4 4 350 200]{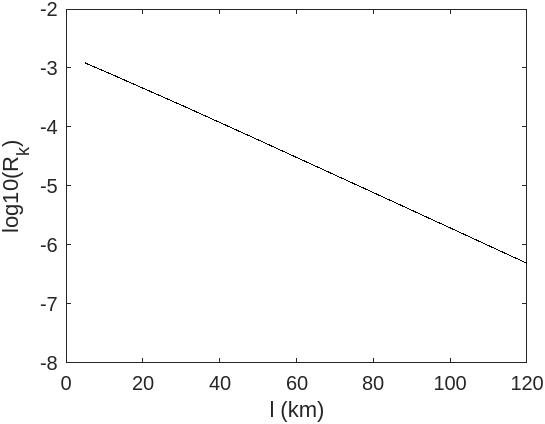}
\caption{Plot of $R_k$ vs. $l$ with $\lambda_s = 0.5$ and $\lambda_d = 0.2$ for 50 km fiber link. Other parameters remain the same as used in Tables 3 and 4.}
\end{figure}

\begin{figure}[ht]
\centering
\includegraphics[width= 4in, viewport=4 4 350 200] {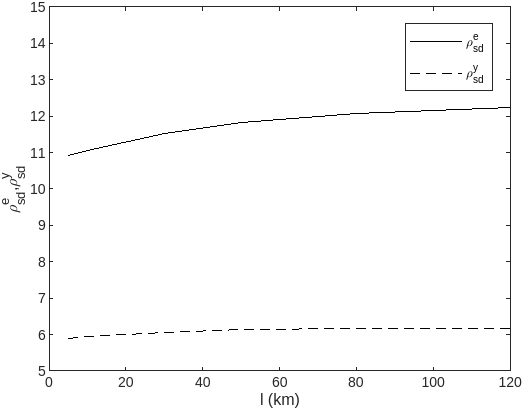}
\caption{Plots of $\rho^e_{sd}$ and $\rho^y_{sd}$ vs. $l$ with $\lambda_s = 0.5$ and $\lambda_d = 0.2$ for 50 km fiber link. Other parameters remain the same as used in Tables 3 and 4.}
\end{figure}

It therefore appears that, the above results obtained from the proposed model are expected to provide highly insightful design guidelines for the QKD protocol under consideration. Furthermore, for the present protocol one can also try different combinations of $\lambda_s$ and $\lambda_d$ for improvement of performance in respect of key generation rate, more so for longer link lengths, while assuring the desired values for $\rho^y_{sd}$ and $\rho^{e}_{sd}$. The proposed model, having been developed from the basic physics of the transmission phenomena, can be readily adapted for various other QKD protocols that have been and are being explored currently by the researchers in this area.   

\section{Conclusion}

This paper presents a novel event-by-event impairment enumeration model to study the performance of a QKD scheme employing BB84-DP-PNS protocol in the presence of potential impairment phenomena in its quantum communication link. With a brief overview of the  the overall system, the deleterious effects of impairments are taken into account in the formulation of the proposed model, including the progressive photon losses in the optical fiber links, Poissonian emission statistics of optical sources, polarization mismatch between the transmitting and receiving nodes, and the losses in the number of bits available for key formation during the photodetection, sifting, and error correction stages in Bob's receiver. 

The results obtained from the proposed model indicate that, the impact of impairments can significantly influence the performance of QKD schemes, and there exists indeed some interplay between the various system parameters that needs to be understood and taken into design considerations to make the protocol secure. In particular, along with the evaluation of the key generation rate, our method helps in choosing appropriate values for the signal and decoy photon counts for a given set of values for $m_s$, $m_d$, $m_v$, and other system parameters, towards ensuring the detectability of Eve's presence from the signal-to-decoy yield ratio observed at Bob. Further, the method ensures that the retrieved two-photon signal pulses by Eve get reasonably cluttered by her two-photon decoy pulses, thus making it harder for Eve to extract the secret key, which in turn makes the task of PA algorithm easier and more effective. Furthermore, with the unique perspective of the entire system, achieved through visualizing all the impairment events along the optical fiber links from Alice to Bob via Eve, the proposed approach proves to be highly useful for looking into and locate/identify the problems if any (e.g., the device parameter settings/choices and possible adjustments therein), and in doing so it provides a realistic design methodology for the QKD scheme under consideration. The proposed enumeration approach, being generic in nature as it is developed from the basic physics of the impairment events, can also be extended readily to analyze and design other QKD schemes, wherever necessary.










\bigskip




\section*{Author Biographies}

\setlength\intextsep{0pt}

\paragraph{}

\noindent \textbf{Debasish Datta} (Life Senior Member, IEEE) received his B.Tech. degree in 1973 from the Institute of Radio Physics and Electronics, Science College, Calcutta University, and M.Tech. and Ph.D. degrees from Indian Institute of Technology (IIT) Kharagpur in 1976 and 1986, respectively. He served in the faculty of IIT Kharagpur for about thirty-seven years and retired in June 2017. In the early phase of his career after completing M.Tech., he worked in the electronics and telecommunication industry for about three years. At IIT Kharagpur, he served as the Chairman of G.S. Sanyal School of Telecommunications and as the Head of the Department of Electronics and Electrical Communication Engineering. 

From IIT Kharagpur, he visited Stanford University during 1992-1993, University of California, Davis, during 1997-1999, Chonbuk National University, South Korea, during 2003-2004, and University of Malaya, Kuala Lumpur during 2013-2014 to work on optical communications and networking. After retirement he has been engaged in lecturing and mentoring researchers in the area of optical networks as visiting professor at various Indian universities. Prof. Datta’s group developed the first WDM transmission system for voice, video and data transmission in India, sponsored by the Department of Electronics, Government of India. He served as Guest Editor for the IEEE JSAC for January-2002 Special Issue on WDM-based Network Architectures. Later he served as Editor for the IEEE Communication Surveys and Tutorials and Elsevier Journal of Optical Switching and Networking. He also served as Track Chair for Core Telecommunication Networks in the Technical Program Committee of IEEE ANTS 2009, as Technical Program Committee Co-Chair in IEEE ANTS 2012 and as General Co-Chair in IEEE ANTS 2015. His current research interests include elastic optical networks, optically-augmented datacenters, and quantum communication.  
\vspace{-20mm}

\end{document}